\documentclass[preprint]{aastex}

\def\beq{\begin{equation}}
\def\eeq{\end{equation}}

\def\cc{{\rm cm}^{-3}}
\def\g{\gamma}
\def\loss{{\rm loss}}
\def\Coul{{\rm Coul}}
\def\sy{{\rm sych}}
\def\scat{{\rm scat}}
\def\esc{{\rm esc}}
\def\eff{{\rm eff}}
\def\tr{{\rm cross}}
\def\ic{{\rm IC}}
\def\br{{\rm brem}}
\def\ph{{\rm ph}}
\def\crit{{\rm cr}}
\def\ac{{\rm ac}}
\def\min{{\rm min}}

\def\mug{\mu{\rm G}}
\def\ln{{\rm ln}}
\def\Clog{\ln\Lambda}

\def\cc{{\rm cm}^{-3}}

\slugcomment{Submitted to \apj, Jan. 2001, Revised March 2001}

\shorttitle{Electron Acceleration  and Non Thermal Emission in Clusters}
\shortauthors{Petrosian}

\begin{document}

\title {On The Non Thermal Emission and Acceleration of Electrons in Coma and 
Other Clusters of Galaxies}

\author{Vah\'{e} Petrosian\altaffilmark{1}}

\affil{Center for Space Science and Astrophysics, Stanford University, 
Stanford, CA 94305}

\email{vahe@astronomy.stanford.edu}

\altaffiltext{1}{Astronomy Program and Departments of Physics and Applied 
Physics}

\begin{abstract}

Some clusters of galaxies in addition to thermal bremsstrahlung (TB), emit
detectable diffuse radiation from the intercluster medium (ICM) at radio, EUV
and hard x-ray (HXR) ranges.  The radio radiation must be due to synchrotron by
relativistic electrons, and the inverse Compton (IC) scattering by the cosmic
microwave background radiation of the same electrons is the most natural source
for the HXR and perhaps the EUV emissions.  However, simple estimates give a
weaker magnetic field than that suggested by Faraday rotation measurements.
Consequently, non-thermal bremsstrahlung (NTB) and TB have also been suggested 
as sources of these emissions.

We show that NTB cannot be the source of the HXRs (except for a short period)
and that the difficulty with that the low magnetic field in the IC model is
alleviated if the effects of observational selection bias, non isotropic pitch
angle distribution and spectral breaks in the energy distribution of the
relativistic electrons are taken into account.  From these consideration and the
strength of the EUV emission, we derive a spectrum for the radiating electrons
and discuss possible acceleration scenarios for its productions.

We show that continuous and in situ acceleration in the ICM of the background
thermal electrons is difficult and requires unreasonably high energy input.
Similarly acceleration of injected relativistic electrons, say by galaxies,
seems unreasonable because it will give rise to a much flatter spectrum of
electrons than required, unless a large fraction of energy input is carried away
by electrons escaping the ICM, in which case one obtains EUV and HXR emissions
extending well beyond the boundaries of the diffuse radio source.  A continuous
emission by a cooling spectrum resulting from interaction with ICM of electrons
accelerated elsewhere also suffers from similar shortcomings.  The most likely
scenario appears to be an episodic injection-acceleration model, whereby one
obtains a time dependent spectrum that for certain phases of its evolution
satisfies all the requirements.

{\it Subject headings}: Galaxies: Clusters: General--Galaxies: Clusters: 
Individual--X-Rays--Magnetic Fields--Acceleration of Particles.
\end{abstract}

\section{INTRODUCTION}

The most prominent radiation from the intercluster medium (ICM) of clusters of
galaxies is the thermal bremssstrahlung (TB) or free-free emission in the soft
X-ray (2 to 10 keV, SXR) region which can reach a luminosity $L_{SXR} \sim
10^{45}$ erg/s and implies gas temperatures of $T \sim 10^8$K and emission
measures of $EM \sim 10^{68} \cc$ (density $n \sim 10^{-3} \cc$, radius $R \sim
1$ Mpc).  For Coma cluster $L_{SXR} \simeq 5\times 10^{44},\, kT=8.2$ keV.
There is, however, a growing evidence for a significant nonthermal activity in
some clusters.  The first of these to be discovered in just a few clusters,
notably in the Coma cluster (for the most recent observations see Giovannini \&
Feretti 2000) was the diffused (so-called halo) radio emission of luminosity
$L_R \sim 10^{41}$ erg/s in the frequency range $30 {\rm MHz} < \nu < 4$GHz
whose source must be synchrotron radiation by relativistic electrons.  The range
and distributions of the Lorentz factor $\g$ of these electrons and their total
energy depends on the strength, geometry and distribution of the magnetic field
$B$.  The field is measured by Faraday rotation to be a few microGauss ($\mug$)
in some clusters which would require electrons with $\g > 10^3$.  The exact
source of these electrons is still a matter of considerable debate.  For a
review see Eilek (1999), Giovannini et al.  (1993), Kim et al.  (1990), and
references there to earlier works.  More recently, radiation (most likely
nonthermal in origin) have been discovered in form of excess flux at low and
high ends of the thermal radiation in several clusters.  {\it The Extreme
Ultraviolet Explorer} (EUVE), with the passband of 69 to 245 eV, has detected
excess radiation from Coma (Lieu et al.  1996a).  The luminosity of this soft
excess radiation in the 0.07 to 0.4 keV range is $L_{EUV} \sim 2\times 10^{43}$
erg/s (Lieu et al.  1999).  Similarly, {\it Beppo}SAX and {\it Rossi X-ray
Timing Explorer} (RXTE) have detected excess hard X-ray (HXR) radiation in the
20 to 80 keV range from Coma (Fusco-Femiano et al.  1999, Rephaeli et al.
1999), A2256 (Fusco-Femiano et al.2000), and possibly A2199 (Kaastra et al.
1999).  The luminosity of Coma in this range is $L_{HXR} \sim 4\times 10^{43}$
erg/s.  The EGRET instrument on board the {\it Compton Gamma-Ray Observatory}
(CGRO) set an upper limit of $L_{\g -Ray} \lesssim 10^{43}$ erg/s above 100 MeV
(Sreekumar et al.  1996).  These observations are summarized in Table 1 and
assume a Hubble constant of 60 km/(s Mpc).

Initially the excess EUV radiation was explained in terms of one or more cooler
thermal components (Lieu et al.  1996a and 1996b, Mittaz et al.  1998) but soon
after, a nonthermal process, namely the inverse Compton scattering (IC) by the
cosmic microwave background (CMB) radiation of electrons with energies about ten
times smaller than those responsible for the radio emission was propsed as the
source of the EUV radiation by many authors (Hwang 1997, En\ss lin \& Biermann
1998, Bowyer \& Bergh\"ofer 1998, Lieu et al.  1999).  The initial
interpretation of the HXR excess was also based on the IC model (Fusco-Femiano
et al.  1999, Sarazin \& Lieu 1998).  This seems to be a natural explanation
since electrons with energies very similar to those producing the synchrotron
emission are required.  In fact, long before these discoveries strong upper
limit on the nonthermal X-ray emission were set based on the IC model and the
radio observations (Schlickeiser et al.  1987).

There is, however, a major difficulty with both of these interpretations (Bowyer
\& Bergh\"{o}fer 1998, Fusco-Femiano et al.  1999, Rephaeli et al.  1999).  This
is due to the simple fact that the ratio of the IC to synchrotron luminosities
is equal to the ratio of the CMB to magnetic field energy densities which for
$T_{CMB}=3$K is $15/(B/\mu{\rm G})^2$.  The observed ratio of HXR to radio
luminosities of about $4\times 10^2$ implies a field strength $B< 0.2 \mu$G
which is much smaller than $B$ values of several $\mu$G deduced from Faraday
rotation (Eilek 1999) and equipartition of magnetic and relativistic particle
energies.  Comparison of the EUV and radio fluxes can also set a limit on the
magnetic field but here the limit is somewhat higher ($B\lesssim 1 \mug$; Hwang
1997, En\ss lin \& Biermann 1998) and less reliable because it is sensitive to
the uncertain extrapolation of the electron spectrum over a decade (Bowyer \&
Bergh\"{o}fer 1998).  Because of this discrepancy, several workers have proposed
nonthermal bremsstrahlung as the source of the observed HXRs (En\ss lin et al.
1999, Sarazin \& Kempner 2000, Blasi 2000a).  However, this explanation also
suffers from a major flaw because it requires a large input of energy in the ICM
whose consequences have not been detected.  This flaw is based on the simple
fact that bremsstrahlung is an inefficient mechanism.

In the next section we describe some details of the characteristics of the ICM
plasma and the constraints they put on the models.  In \S 3 we discuss the
emission process and in \S 4 the related aspect of the particle acceleration.  A
brief summary is presented in \S 5.

\begin{center}
\center{{\Large {\bf Table 1}}}
\center{\bf SOME COMA CLUSTER OBSERVATIONS}

\begin{tabular} {|l|cccc|}
\hline \hline
Radiation & Radio & EUV &SXR &HXR 
\\ 
\hline
Range & $0.03-4$ GHz & $0.07-0.4$ keV & $2-10$ keV & $20-80$ keV  
\\
Spectrum &  Broken Power Law & Uncertain & Exponential & Power Law
\\     
Index (or Temp.) & $\sim 1$ and  $\sim 2$  & $\sim 3/2$ &  
$kT=8.5^{+0.6}_{-0.5}$keV 
&  0.7 to $6^a$
\\     & Exponent. Tail &  $kT\sim 2$keV       &  $k=7.51\pm 0.18$ keV & 
$2.35\pm 
0.45^b$
\\
Lum. (erg/s) &  $10^{41}$      &$2\times 10^{43}$ & $5\times 10^{44}$ & $4\times 
10^{43}$ 
\\
Mechanism & Synchrotron & IC (TB) & TB& IC (NTB)
\\
\hline \hline
\end{tabular}
\end{center}\vspace{-0.2in}
\hspace{0.15in}a) Fusco-Femiano et al. (1999)
\hspace{0.15in}b) Rephaeli et al. (1999)

\section{GENERAL DESCRIPTION OF THE PROBLEM}

In order to illustrate the difficulties faced in the above models, in Figure 1
we show the energy loss timescales, $\tau_{\loss} = -E/{\dot E}_\loss$, as a
function of particle kinetic energy for all the relevant processes in this
problem.  Here and in what follows unless explicitly expressed all energies and
loss rates will be in units of rest mass energy of electron, $m_ec^2$, so that
the Lorentz factor $\g=E+1=(1-\beta^{2})^{-1/2}$.  We use the following
expressions for the loss rates.  
\beq\label{ic} 
{\dot E}_\ic = (32\pi/9)r_0^2c\beta^2\g^2u_{\ph} = 2.04\times
10^{-20}\beta^2\g^2(T_{CMB}/3)^4\,\, {\rm s}^{-1}, 
\eeq 
\beq\label{sy} 
{\dot E}_\sy = (4/9)r_0^2c\beta^2\g^2B^2 = 1.32\times
10^{-21}\beta^2\g^2(B/\mug)^2\,\, {\rm s}^{-1}, 
\eeq 
\beq\label{Coul} 
{\dot E}_\Coul = 4\pi r_0^2cn\Clog/\beta = 1.20\times 
10^{-15}(n/10^{-3}\cc)/\beta\,\, {\rm s}^{-1}, 
\eeq 
\beq\label{br} 
{\dot E}_\br = (16/3)\alpha r_0^2cn\beta\g\chi(E)= 9.29\times 10^{-20}(n/10^{-3} 
\cc)\beta\g\chi(E)\,\, {\rm s}^{-1}, 
\eeq
where $\alpha$ is the fine structure constant, $u_\ph$ is the soft
photon energy density.  The IC losses are evaluated assuming the
CMB with temperature of 3 K as the source of the soft photons.  The synchrotron
losses are evaluated for an isotropic pitch angle distribution and the Coulomb
logarithm $\Clog$ is set to 40, a representive value for the ICM conditions.
For the Coulomb and bremsstrahlung losses we assume presence of 10\% (by number)
of fully ionized helium.  The bremsstrahlung rate also depends on the complex
function $\chi(E)$ which is equal to one in the nonrelativistic limit and equals
$\threequarters[\ln (2E) - \onethird]$ at extreme relativistic energies.  (For
$E \gg\alpha^{-1}$ the slow varying term in the square brackets tends to the
constant value of about 5.)  The bremsstrahlung rate used in Figure 1 is 
calculated
from the more exact expression given by the Formula 4BN of Koch \& Motz (1959).

\begin{figure}[htbp]
\leavevmode\centering
\epsscale{1.0}
\plotone{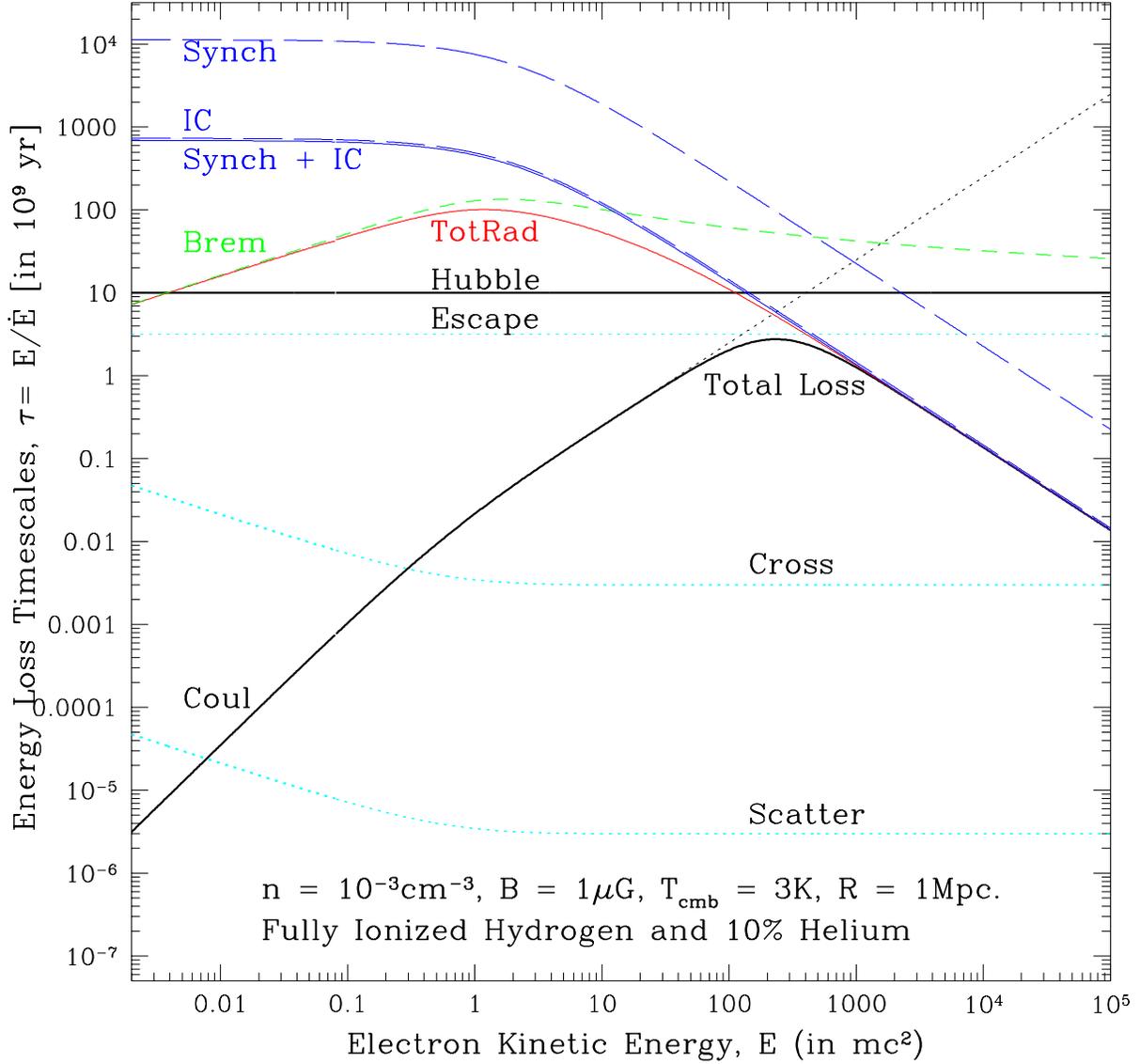}
\caption{The energy loss timescales vs energy for the four relevant 
interactions 
of electrons for typical ICM conditions. The three solid lines, from top to 
bottom, are for  the combined synchrotron and inverse Compton, the three 
radiative processes, and all losses,  respectively. The dotted lines show 
the average crossing time $T_\tr \sim R/(c\beta)$ across a region of size 
$R\sim 1$Mpc, the scattering time $\tau_\scat \sim \lambda_\scat/(c\beta)$ for 
a 
constant scattering mean free path $\lambda_\scat$, and the escape time $T_\esc 
\sim T_\tr^2/\tau_\scat$. Note that all loss times are shorter than the Hubble 
time 
(heavy solid line for Hubble constant of 100 km s$^{-1}$ Mpc$^{-1}$), and much 
longer than the electron crossing time except at low ($E<200$ keV) and very high 
($E>200$ GeV) energies.}
\end{figure}

Several immediate conclusions can be drawn from the above figure.
 
The lifetimes of electrons with energies in the range $200{\rm keV} \leq E \leq 
200{\rm
GeV}$ are longer than the free crossing time of the electrons
across the cluster (or the 'mean free paths', $\lambda_\loss = c\beta \tau_\loss
\lesssim 1$ Gpc, are much larger than the size $R \sim 1 $ Mpc of the cluster).
Therefore, these electrons, if unhindered, {\it e.g.}  by chaotic magnetic 
fields
or other scattering agents, will escape the cluster before losing most of their
energy and while in the cluster they will radiate what is commonly referred to 
as
a {\it thin target} spectrum.  The escaping electrons will radiate most of their
energy outside the cluster, presumably by IC scattering of the CMB photons.
This will disagree with the observations and will require a higher rate of
energy input than for electrons outside the above energy range which lose all
their energy before escape and develop a so-called {\it cooling spectrum} and
giving rise to a {\it thick target} photon spectrum.  Therefore, if all
electrons were to lose all their energy in the cluster, they must be trapped
efficiently so that they traverse a Gpc in the ICM.  This can come about by a
thousand reversals of the magnetic field lines or a million random scattering of
the electrons.  Hence,

\noindent
{\it we require the presence of scattering agents (e.g.  plasma turbulence) with
a scattering mean free path $\lambda_\scat\sim 1$ pc, or a chaotic field
structure with a scale $B/\Delta B$ between 1 kpc to 1 pc}.

\noindent 
In such a case, the electrons with energies away from the peak energy $E_p$ of
the total loss time curve at about 100 MeV diffuse through only a distance of
$\lambda_\eff\sim (\lambda_\loss \lambda_\scat)^{1/2}\ll R$ before they lose 
most of their energy.  This means that for production of a smooth diffuse 
radiation
throughout the cluster

\noindent
{\it we need in situ acceleration of the electrons throughout the ICM} 

\noindent and cannot rely on injection of accelerated electron into the ICM from
a single source or sources separated by a distance $\gg \lambda_{\rm eff}$.
This last condition is required for electrons with $E<200$ keV and $E>200$ GeV
even in the absence of scattering agents.  (Berezinky, Blasi \& Ptuskin, 1997, 
address the issue of the confinement of the non thermal particles using the 
scattering mean free path given by Schlickeiser et al. , 1987. However, their 
discussion is applicable only to protons, and other ions, because they ignore 
radiative and other losses. As evident from Figure 1 this cannot be the case for 
electrons.) We will return to these requirements
in \S 4 dealing with the acceleration process.

On the other hand, because the lifetimes at all energies are much shorter than 
the age of the universe, then unless the 
observed nonthermal radiations are short lived transient phenomena,

\noindent
{\it the acceleration of the nonthermal electrons must be continuous over the 
lifetime of the clusters}, 

Finally, we note that for $B<1\mug$, the synchrotron process has little
influence on the dynamics (acceleration and cooling) of the nonthermal
electrons.  It acts only as the radio emission process.

\section{EMISSION PROCESSES}

In this section we describe the difficulties faced in some of the proposed 
radiation mechanisms and derive a spectrum for the nonthermal electrons.

\subsection{NonThermal Bremsstrahlung Emission}

En\ss lin et al.  (1999) were the first to propose this emission (NTB) as the 
source
of the observed HXR flux from clusters, whereby electrons of comparable or
slightly larger energies produce the 20 to 80 keV radiation.  Sarazin \&
Kempner (2000) evaluated bremsstrahlung spectra using various accelerated
electron spectra and  detailed bremsstrahlung cross sections.  Blasi (2000)
gives a combined description of the stochastic acceleration and bremsstrahlung
radiation.  However, all these works ignore the huge energy problem associated
with this model.  As is evident from Figure 1, the main difficulty of this model
is the inefficiency of the bremsstrahlung process compared to the collisional
losses for $E<1$ GeV and relative to IC losses for $E>10$ MeV.  In particular,
for the energy range of interest here (20 to $\sim 1000$ keV) the ratio of the
bremsstarhlung to Coulomb loss rates is less than $10^3$.  As shown by Petrosian
(1973) the yield of the bremsstrahlung photon is a well defined quantity
independent of many unknowns of the models.  For a non relativistic electron of
initial energy $E_{in}$ that loses all its energy ({\it thick target case}) this
yield is $Y_{\br}=(4/3\pi)(\alpha/\Clog)E_{in}=7.7\times 10^{-5}E_{in}$.  (This
yield is larger by a factor of two for electrons losing a small fraction of
their energy in the source region, {\it thin target case}).  For a power law
distribution of electrons ($N(E)\propto E^{-p}$, for $E>E_{in}$), the above
expression is modified by a factor of order unity: The yield of electrons with 
energies between $E_{in}$ and $E_f$ will depend on $p$. For $p\sim 3.5$ required 
by this model this factor is 1.3 (see eq. [31] in Petrosian 1973, where $\delta 
+ 1 = p$).  This
expression is also valid for relativistic energies within a factor of the order
of $\ln E_{in}$ as indicated by the slow decline of $\tau_\br$ curve in Figure 1
at high energies.

A yield of $Y_{\br}<3\times 10^{-6}$ in the 20 to 80 keV range means that, 
independent
of most details of the acceleration or emission model, a large amount of energy
($L_{\rm in}=L_{HXR}/Y_{\br} \sim 10^{49}$ ergs/s) is fed into the background
plasma.  If the ICM plasma were to cool only radiatively (free free emission),  
at the very slow rate of $L_{\rm ff}= 1.45\times
10^{-23}{\rm ergs/s}(T/10^8 {\rm K})^{1/2} (EM/10^{68} \cc )$, then such an 
input of energy will increase its temperature at the rate of $dT/dt=L_{in}/3Nk 
\gtrsim 10^{-7}$ K/s. As a result the ICM temperature will exceed $10^8$ K after 
a short time of $3\times 10^7$ yr and will exceed $10^{10}$ K in a Hubble time! 
This, of course, is not acceptable because it
will evaporate the ICM plasma into the general intergalactic medium. Either only 
one 
part in $10^{4\, {\rm or}\, 5}$ of the observed HXR flux is due to the NTB 
process or the NTB emission phase at the observed rate is a short lived 
phenomena. Blasi
(2000a) finds that his acceleration model indeed requires a similar (though 
somewhat smaller) rate of input of
energy into the turbulence needed for acceleration, and that the duration of NTB 
emission satisfying the observation is around several hundred million years.

The situation is very similar in what one may call the {\it inverse
bremsstrahlung} model, whereby accelerated protons interacting with the
background thermal electrons produce the HXRs.  In the rest frame of an
accelerated proton of energy $E_p=(m_p/m_e)E_e$ the process is identical to that
of bremsstrahlung by accelerated electrons of energy $E_e$.  Thus, HXRs of
energy 20 to 200 keV can be produced by nonthermal protons of energy 40 to 1000
GeV.  However, here again, most of the proton energy will go into heating the
electrons by inelastic Coulomb or Rutherford scattering.  In addition, the
higher energy nonthermal protons may lose some of their energy to $\pi^o$
production which decay into 50 to 100 MeV gamma-rays.

The presence of thermal SXRs and nonthermal HXRs (also nonthermal synchrotron
radio) emission in the clusters is very similar to that observed in solar
flares.  Except in solar (and most likely in other stellar) flares the SXR flux
is $10^5$ to $10^6$ times larger than the HXR flux in agreement with the above
yields.  In analogy to flares one may consider acceleration of electrons is
taking place in high density magnetic loops associated with the disks or halos
of, say a thousand galaxies, each receiving $10^{45}$ ergs/s.  The current
observations do not have the spatial resolution to distinguish the ICM emission
from that of many galaxies.  Since the radiative equilibrium temperature
$T\propto (L_{\rm in}/(Y_\br EM))^2$, a lower temperature of about $10^9$K will
result for a $L_{\rm in}\sim 10^{45}$ ergs/s, density $n\sim 0.1$ and region of
size $L\sim 30$ kpc. (Actually for this size scale conduction losses $L_{\rm
cond}\sim 2\times 10^{45}(T/10^8{\rm K})^{3.5}(L/30{\rm kpc})$ become comparable
and exceed the radiative losses for $T\geq 10^8{\rm K}$ so that the temperature
never exceeds this latter value.) In any case such hot galactic plasmas will 
evaporate into
the ICM and may be the source of the hot SXR emitting gas.  Only a small
fraction ($<0.01$\%) of the energy can go into the ICM plasma.  Most of it must
be dissipated in the galaxies.  It is not obvious how the effects of such an
energy input which is much larger than that from stellar sources (stellar winds,
supernovae and other explosions) can be hidden.

Therefore, we conclude that the main objection to the NTB
emission is very robust; it is essentially determined by the values of the fine
structure constant and the Coulomb logarithm and very difficult to circumvent.
This leads to the conclusion that for all three clusters, Coma, A2256 and A2199,

\noindent {\it the NTB emission from ICM as source of the
observed HXRs is not tenable, unless it is a short-lived ($<10^8$ yr) 
phenomenon.}

A corollary of this is that one can put a strong constraint on the spectrum and
energy density of the nonthermal electrons below the peak energy $E_p\sim
100$MeV where the elastic Coulomb collision loses are larger than the radiative
losses.  As we will show below, the spectral distribution of the electrons below 
this
energy must be flatter than $E^{-1/2}$ or there must exist a sharp cut off below
several MeV.

\subsection{Inverse Compton Emission}

As can be seen from Figure 1, for typical ICM conditions, the IC emission 
exceeds
bremsstrahlung for $E>10$ MeV.  However, as already pointed out by many of the
authors cited in the introduction this model also suffers from the inefficiency
of the IC radiation relative to the synchrotron radiation.  The relative flux of
these two radiations depends on the CMB photon density, which is known
accurately, and on the value of the magnetic field which is not so well known.
From equations (\ref{ic}) and (\ref{sy}) the ratio of these fluxes can be
obtained to be roughly equal to
\beq\label{ratio}
{\dot E}_\ic/{\dot E}_\sy = 19.8(T_{CMB}/2.8{\rm K})^4(\mug/B)^2.
\eeq
The most reliable measures of ICM $B$ field come from the Faraday rotation of
the background radio sources.  In the cores of several well studied clusters
values of several $\mug$ have been derived (Eilek 1999).  Furthermore, These
refer to the net line of sight component so that for a chaotic field the actual
value could be even larger.  This is in apparent contradiction with the value
$B\sim 0.2\mug$ one deduces from the observed ratio of the HXR to radio fluxes,
which is about 500.  Actually, the observed Faraday rotation in the Coma cluster
gives a value of $B\sim 0.3\mug$ for an ordered magnetic field and a larger
value of $B\sim 2\mug$ if the field is chaotic on the scale of several tens of
kpc (Kim et al.  1990, Ferretti et al. 1995).  Estimates based on the assumption 
of energy
equipartition between nonthermal electrons, protons and magnetic field give $B
\sim 1\mug$.  The radio properties are somewhat different for A2256 but the same
kind of discrepancy seem to be present for this cluster as well (Fusco-Femiano
et al.  2000).  However, there is no Faraday rotation data for this cluster so
the argument against the IC model is not as strong here.  The detection of HXR
in A2199 is generally considered as marginal, which combined with the absence of
a detectable halo radio source or Faraday rotation makes conclusions based on
this cluster less reliable.

However, in comparison with the insurmountable difficulty of the nonthermal
bremsstrahlung model, there are possible ways to avoid the problems of the IC
model.  Below we describe several effects that alleviate these problems.  We
will use Coma cluster for quantitative discussion.

\subsubsection{Selection Effects}

The low value of the $B$ field in clusters with observed HXR emission can be
simply an observational selection effect.  For a given radio flux of the halo
source, and independent of any equipartition argument, clusters with lower $B$
values will have the stronger IC flux and, therefore, will be more readily
detected by {\it Beppo}SAX and RXTE (or EUVE if electron spectrum extends to
lower energies).  Unfortunately the numbers of known clusters with either (or
both) a halo radio source and HXR emission are two small to make any reliable
quantitative estimates of the effects of this selection bias.

A related and similar effect can arise if the distrubutions of the magnetic 
field and relativistic electrons are inhomogeneous and anticorrelated. In this 
case the radio and IC emissions will come mainly from weak field regions while 
the Farady rotation is determined by the average field. Even in the absence of 
such an anticorrelation, there are other subtle effects arising from spatial 
inhomogeneities that can give rise to 
a discrepancy between the magnetic field strengths based on the IC emission and 
the Faraday rotation measure (Goldshmidt \& Rephaeli 1993). Explorations of 
spatially inhomogenous models is beyond the scope of this paper.

\subsubsection{Complex Electron Spectra}

The estimate of ratio of IC to synchrotron emission based on equation
(\ref{ratio}) is for a monoenergetic electron.  For a spectrum of accelerated
electrons this relation is somewhat more complex.  However, for a power law
distribution of accelerated electrons with index $p$ one obtains similar
constrains on the value of the magnetic field using the observed HXR and radio
fluxes.  Using the well known expression for the spectra of IC and synchrotron
emissions it can be shown that the ratio of the HXR photon flux (in units of
ph/(s cm$^2$ keV) at photon energy $\epsilon$ to the radio flux (in Jy units) at
frequency $\nu$ is
\beq\label{fratio}
R={f_\ic(\epsilon) \over f_\sy(\nu)}=1.86\times 10^{-8} \left({{\rm photons} 
\over {{\rm s \,cm}^2\, {\rm keV \,Jy}}}\right)\left({\epsilon \over {20{\rm 
keV}}}\right)^{-\alpha}\left({\nu \over {\rm 
GHz}}\right)^{\alpha-1}\left({T_{CMB} \over {2.8{\rm 
K}}}\right)^{\alpha+2}\left({B_\perp \over \mug}\right)^{-\alpha}g(p),
\eeq
where $\alpha=(p+1)/2$ is the IC photon spectral index and $g(p)$ is a
complicated function of index $p$ which is equal to 11.0, 41.2, 181 and 755 for
$p=2, 3, 4$ and 5, respectively (see eqs.  [6.36] and [7.29], Rybicki \&
Lightman 1979).  In this range of $p$ a good approximation to use is
$g(p)=e^{(1.42p-0.51)}$.  Using this approximation it can be shown that
\beq\label{Best}
(B_\perp/\mug)=(20 {\rm keV}/\epsilon)(\nu/{\rm GHz})^{{p-1}\over {p+1}}{\rm   
exp}\{2.84(p+r)/(p+1)\},
\eeq
\beq\label{rfactor}
r=0.7\ln (R_{obs}(\epsilon, \nu)/1.11\times 10^{-8}),
\eeq
where $R_{obs}(\epsilon, \nu)$ is the observed ratio of the fluxes.  Using the
observed flux ratios from Coma at several values of $\epsilon$ and $\nu$ we find
that for $p=3$ the field strength is $B_\perp=0.18\mug$.  The required value of
$B_\perp$ increases with $p$ monotonically but slowly.  For example, for $p=5$
we find $B_\perp=0.8 \,{\rm to}\, 0.3\mug$ depending on the values of $\epsilon$
and $\nu$.
This indicates that magnetic fields of about $1\mug$ may be possible if the
electron spectrum steepens at some energy just below that needed for production
of HXRs.  

The energy range of electron needed for production of observed HXR (20
to 80 keV), EUV (0.07 to 0.4 keV) and radio (0.03 to 3 GHz) are
\begin{eqnarray}
\label{rHXR}
0.53 & <(E_{HXR}/10^4)< & 1.2,\\
\label{rEUV}
0.3 & <(E_{EUV}/10^3)< & 0.75,\\ 
\label{rradio}
0.4(\mug/B_\perp)^{1/2} & <(E_{rad}/10^4)< & 4(\mug/B_\perp)^{1/2}. 
\end{eqnarray}
Note that radio waves with $\nu>0.35$GHz are emitted by electrons above the
range needed for the other emissions.  Thus, a steepening of the accelerated
electron spectrum at $E=E_\crit\sim 10^4$ will reduce the radio flux and allow a
higher magnetic field.  For example, if the spectral index of the electrons
changes from 3 to 5 at $E_\crit$ (as in the Rephaeli 1979 model), then following
equations (\ref{fratio}) and (\ref{Best}), it can be shown that we need
$B_\perp\simeq 0.5\mug(E_\crit/10^4)^{-2}$.  Even higher magnetic fields will be
allowed if the electron spectrum cuts off exponentially, as is the case for some
acceleration models described by Schlickeiser et al.  (1987, see also below ).  
For
$N(E)=N_0(E/E_\crit)^{-p}{\exp\{E/E_\crit\}}$, the radio flux at high
frequencies is reduced approximately by a factor of
$(p-1/3)E_{p+2/3}[(\nu/\nu_\crit)^{1/2}]$, where $\nu_\crit=0.42{\rm
GHz}(B_\perp/\mug)(E_\crit/10^4)^2$ and $E_n(x)$ is the exponential integral
function.  (This result is obtained by approximating the monoenergetic
synchrotron spectrum as $\eta(\nu, E)=A(\nu/\nu_c)^{1/3}$ for $\nu\leq
\nu_c=3E^2\nu_B/2$).  Schlickeiser et al.  (1987) show that a power law spectrum
with an exponential cut off at $\nu_\crit=0.15$GHz provides a much better fit
than a single or a double power law model.  With this cut off frequency the
application of the above correction factor yields $B_\perp\simeq 1.7\mug$ and
$B_\perp\simeq 1.1\mug$ for $p=3$ and 4, respectively.  These higher field
strengths are in better agreement with the Faraday rotation measures quoted
above.

We will return to these consideration in \S {\bf 3.3} and {\bf 3.4} and show
that these requirements set further constraints on the acceleration mechanism.

\subsubsection{Anisotropic Pitch Angle Distribution}

The gyroradius of the nonthermal electrons $r_g=2\pi c\beta\gamma/\nu_{B_\perp}
\sim 10^{11}{\rm cm}\beta\gamma(\mug/B_\perp)$ is much smaller than all other
relevant scales in clusters.  Therefore, the electrons are attached to the field
lines and their distribution can be described by a gyrophase averaged
distribution $N(E, \psi)$ as a function of energy and pitch angle $\psi$.  In
the above discussion we have implicitly assumed an isotopic pitch angle
distribution.  Anisotopies can modify some of the results quoted above.  Note
that all values of magnetic field are quoted in terms of $B_\perp$.  For an
ordered field and isotropic pitch angle distribution the synchrotron emissivity
is related to the component of the field perpendicular to the line of sight.
However, as stated above the magnetic field, even though ordered on the scales
comparable to and larger than $r_g$, must be chaotic on kpc scale.  The 
emissivity averaged over
scales larger than one kpc will be isotropic independent of any anisotropies in
the monoenergetic emissivities and the pitch angle distribution.  However the
overall intensity will depend on the pitch angle distribution.  In this case
$B_\perp=B \sin \psi$.  The synchrotron emissivity at a given frequency is
proportional to $B_\perp^{(p+1)/2}$ so that the above field values must be
corrected by the value of $(\sin \psi)^{(p+1)/2}$ averaged over the pitch angel
distribution in the range $0<\psi<\pi /2$.  If the distribution is isotropic the
average value of this quantity is 2/3 and 8/15 for $p=3$ and 7, respectively, so
that the actual values of magnetic field will be 1.5 to 2 times larger than
those quoted above and could be as high as $B=3\mug$.

Even higher fields will be required if the pitch angle distribution is
anisotropic and is beamed along the field lines.  For a Gaussian pitch angle
distribution of width $\psi_0<1$ the field strengths increase by a factor of
$\psi_0^{-q}$, where the value of $q$ depends on several factors but is greater
than one and could be as high as a few.  The spectral shape also deviates from
the usual power law with index $\alpha_\sy= (p -1)/2$ depending on the values of
$\psi_0$ and $\psi_0\g$.  For further details on this see Epstein (1973) and
Epstein \& Petrosian (1973).  Whether the acceleration mechanism will accelerate
$E>10^4$ electrons preferentially along a jumbled field line depends on the 
conditions in the background plasma.  We
will return to this in \S 4 wherre we will argue in favor of the isotropic 
distribution.

\noindent {\it In summary, there does not appear to be an insurmountable
discrepancy between the field strengths required by the IC model for HXRs and 
the
observed values}.

\subsection{EUV Emission and Electron Spectral Index}

If the EUV emission is also produced by the IC process the nonthermal electron
distribution must extend to $<100$ MeV.  Unfortunately the electron spectral
shape in this range is not well determined and we must rely on the extrapolation
of the spectra from $10^4$ MeV range which can lead to a large uncertainty. 

As already alluded to in the previous section there has been considerable
discussion of the radio spectrum and its implication for the electron spectrum.
A single power law fit gives a value of radio spectral index of about
$\alpha_\sy=1.5$ implying an electron spectral index of $p=4$.  However, as
pointed out by Schlickeiser et al.  (1987), broken power laws provide a better
fit.  For example, the fit to the Rephaeli (1979) model yields an spectrum with
index of about 1 which steepens to 2 above 0.6 GHz, implying an electron
spectral indices of $p=3$ and $p_h=5$ respectively below and above the break
energy of $E=1.6\times 10^4(\mug/B_\perp)^{1/2}$.  Even better fit is obtained
for a spectrum with an exponential cut off $\eta(\nu )\propto \nu^{-0.52}e^{(\nu
/0.15{\rm GHz})}$, which means an index of $p=2$.  However, the range of the
acceptable low frequency spectral indices is fairly large.  For the last model
the 90\% confidence range of $\alpha_\sy$ extends from +0.3 to -1.0 implying
$-\infty < p < 3$ (see Schlickeiser et al. 1987).

The HXR spectrum is even more uncertain.  Fusco-Femiano et al.  (1999) give
photon index $0.7 < \alpha < 3.6$ which allows $0.4 < p < 6$.  Rephaeli et al.
(1999) give a similar value but a smaller range of $1.9 < \alpha < 2.8$ so that  
$2.8
< p <4.6$.  The EUV observation when fitted to a power law indicate a photon
spectral inedex in the range 1.3 to 2.0.  If the EUV emission is also due to the
IC process these values of the photon index indicate a low energy
($10^2<E<10^3$) electron index in the range $1.6 < p < 3.0$.  For a summary of
these observations see Table 1.

It therefore appears that a value of $p\lesssim 3$ is consistent with most of
the data.  A value of $p=3$ implies an IC photon spectrum $f(\epsilon) \propto
\epsilon^{-2}$ and equal energy emission per decade.  The ratio of the observed
EUV flux in the 0.07 to 0.4 keV range of $1.5 \times 10^{-11}$  to the HXR (20 
to 80
keV) flux of $2.2\times 10^{-11} {\rm ergs/(s\,} \cc)$ (see Lieu et al. 1999 and 
Fusco-Famiano et al. 1999) would indicate a $p\sim
2.9$, which is also consistent with the above values.

{\it In summary, the EUV, HXR and radio data can be fitted by the IC and
synchrotron emission in a chaotic magnetic field of strength around 1 to 2 
$\mug$, by electrons with the same spectral distribution as that needed for the 
production of the observed radio spectrum via the synchrotron process.}

\subsection{Spectrum of Radiating Electrons}

From the above discussions we can constrain the instantaneous spectrum of the
{\it radiating electrons} as follows.  We will assume an isotropic pitch angle
distribution.

The radio and HXR observations indicate presence of a power law electron
spectrum with an index $p<3$ and sharp (preferably exponential) cut off at
$E>E_\crit\backsimeq 10^4(\mug/B_\perp)^{1/2}$.  If the EUV emission is also due
to IC process the electron spectrum must extend to about 100 MeV with a somewhat
lower spectral index ($p\backsimeq 2.8$).  At this energy about half of the
electron energy is lost through Coulomb collisions and about 10\% is radiated as
bremsstrahlung photons of $\epsilon < 100$ MeV.  Below $E=200(10^{-3}\cc /n)$
the collision losses become dominant and for $E\leq 10(10^{-3}\cc /n)$
bremsstrahlung surpasses all other emissions.  Therefore, there is a limit how
far this spectrum can be extended.  It can easily be shown that if the electron
spectrum is extended below 20 MeV with $p=3$ the collisional heating rate of the
background thermal plasma will exceed the rate of SXR thermal bremsstrahlung
emission rate.  Since there are other sources of heating of the plasma the
electron spectrum must cut off rapidly at this or higher energy.  However, if
the spectral change occurs at a higher energy the cut off does not necessarily
have to be so severe.  For example, a spectral break at $E_\min\sim E_p\simeq
200$ MeV with index $p_l\leq 0.5$ can be extended to very low energies without
violating the heating rate threshold.  Note that in any case the nonthermal
bremsstrahlung emission in the X-ray (20 to 80 keV) as well as gamma-ray
($\gtrsim 10$MeV) range will be negligible compared to the observed HXR flux.
It can reach at most about 10\% of the total losses at 200 MeV.  Also note that
the commonly used spectrum which is a power law in terms of Lorentz factor
$\gamma$ (see {\it e.g.}  Giovannini et al.  2000) has a natural break
(flattening) below 0.5MeV.  Such a spectrum also must flatten much before below
$\gamma\sim 100$ to avoid the violation the above mentioned threshold.

There has been some discussion (see Blasi 2000b, Bykov, Bloemen \& Uvarov 2000, 
Sarazin 2000) of a possible constraint imposed by the observed EGRET upper limit 
of  $\sim 6\times 10^{-12}$ ergs/(cm$^2$ s) above 100 MeV (Sreekumar et al. 
1996), which is about 0.3 of HXR and 0.6 of EUV fluxes. This constraint  is not 
very stringent because the expected bremsstrahlung flux by electrons above this 
energy is about 30 times smaller than their EUV emission or the HXR emission by 
higher energy electrons (see Figure 1).

In summary, the radiating electron spectrum can be  described as
\beq\label{rade}
N(E) = N_0\cases{(E/E_\crit)^{-p}e^{-(E-E_p)/E_\crit} &if  
$E_p<E,\,\,\,\,{\rm for}\,\,\,\, p\lesssim 3$; \cr
(E/E_p)^{-p_l}(E_p/E_\crit)^{-p} &if  $E<E_p,\,\,\,\,{\rm for}\,\,\,\, p_l\leq 
0.5$.\cr}
\eeq
In the next section we discuss the types of acceleration mechanisms and plasma 
conditions that can give rise to such an spectrum. However, the reader is 
reminded that all the results in this section, including the above equation, are 
applicable to Coma and only to other clusters with similar observational 
characteristics.

\section{ELECTRON ACCELERATION}

The above spectrum is not necessarily that of the {\it accelerated electrons}.
It would be in case of thin target emission when only a small fraction of energy
is lost during the radiation process in the ICM, {\it i.e.}  if
$\tau_\loss>T_\tr\sim R/c\beta$.  As discussed in connection to Figure 1, this
would appear to be the case for electrons with 200 keV $< E < 200$ GeV.
However, we face two critical problems if the electrons escape the ICM in a time
scale shorter than their loss times.  The first is that this requires an
unreasonably high amount of  energy for  acceleration of the electrons;
electrons in the relevant energy range radiate less than 1\% of their energy in 
the ICM.  The
second problem is that escaping electrons will continue to produce IC photons
outside the ICM and will give rise to EUV and HXR emission that extends well
beyond the cluster boundary as determined by the TB and radio (halo) emissions.  
This is not what is observed, especially at EUV
energies where the source has been resolved (see Bowyer \& Bergh\"{o}fer 1998).
Consequently, all electrons must be trapped, by chaotic fields or turbulence,
and lose all their energy in the ICM as in a thick target model.  It is
therefore the totality of the acceleration, scattering and loss processes which
determine the spectrum of the radiating electrons.  In this section we discuss
some general features of the acceleration process and the conditions which can
give rise to the spectrum given by equation (\ref{rade}).

\subsection{General Features of Acceleration}

The trapping of the electrons requires that they undergo repeated deflections or
scatterings such that their effective transport time across the cluster, which
we will refer to as the escape time, $T_\esc = T_\tr(R/\lambda_\scat) >
\tau_\loss$, where $\lambda_\scat = c\beta\tau_\scat$ is the scattering (or
deflection) mean free path.  For this to be true for EUV emitting electrons
($E\ga 200$MeV) we need $T_\esc\ga 3\times 10^9$ yr or a $R/\lambda_\scat\ga
10^3$; {\it i.e.}  we need more than million random scatterings.  This implies a
mean scattering time scale $\tau_\scat\la 3\times 10^3$ which is more than
$10^3$ times shorter than the crossing time and much shorter than all other
relevant times (see Fig. 1).

A secondary effect of the repeated scatterings is that the pitch angle
distribution of the electrons will be isotropic.  The short mean free path also
means that HXR and radio emitting electrons traverse distances equal to 1/40 and
1/200th of the cluster radius within their lifetimes.  Consequently, for a
smooth diffuse source the acceleration must be occurring throughout the ICM with
inhomogeneity scale smaller than a few kpc, or the resolution of the observation 
if it is larger. (It should be noted that one can 
impose an {\it ad hoc} energy dependence scattering process with mean free path 
$\lambda_\scat(E)=R^2/\lambda_\loss(E)$ so that the effective range os all 
electrons is $\sim R$ and $T_\esc(E)=\tau_\loss(E)$; see below.)  We are
therefore dealing with essentially a {\bf homogeneous and isotropic} situation
in which case the general Fokker-Planck transport equation describing the
gyrophase and pitch angle averaged spectrum, $f(E,t)$, of the accelerated
electrons is simplified to
\beq\label{KEQ}
\frac{\partial f}{\partial t} = \frac{\partial ^2}{\partial E^2} [D(E) f] - 
\frac{\partial}{\partial E}[(A(E) - |\dot{E}_{L}|)f] -\frac{f}{T_\esc (E)} + 
Q(E,t).
\eeq
Here $D(E)$ and $A(E)$ are the diffusion and systematic acceleration
coefficients, $Q(E,t)$ is a source term, $\dot{E}_{L}$ is sum of the loss terms
given in equations (\ref{ic}) to (\ref{br}), and $T_\esc (E)$ is the escape
time.

For {\bf stochastic acceleration by turbulence}, a second order Fermi 
acceleration
process, $D(E)=\beta^2 D_{pp}$ and $A(E)={D(E) \over E})\left({{d\ln 
D}\over{d\ln E}}+{{2-\g^{-2}}\over{1-\g^{-1}}}\right)$ describe the diffusive 
and systematic accelerations, where $D_{pp}$ is the momentum diffusion 
coefficient.
From these we can define energy diffusion and acceleration times $\tau_{\rm
diff}= E^2/D$ and $\tau_\ac = E/A$.  The escape time is related to the mean
scattering time $\tau_\scat\sim D_{\mu \mu}^{-1}$, where $D_{\mu \mu}$ is the
pitch angle diffusion coefficient; $T_\esc=T_\tr^2D_{\mu \mu}$.  For
relativistic particles in resonant interaction with Alfv\'en or Whistler waves,
for example, $\tau_\ac\sim (\beta/\beta_A)^2\tau_\scat$, where $c\beta_A$ is the
Alfv\'en velocity (for further details see e.g.  Hamilton \& Petrosian 1992).
For ICM conditions $\beta_A\simeq 2.3\times 10^{-4}$, which means
$\tau_\ac\simeq 10^7\tau_\scat$.  For an efficient acceleration we need an
acceleration time which is shorter than both the escape and the energy loss
times.  For the relevant energies this means $\tau_\ac\la 10^8$ yr,
$\tau_\scat\la 10$ yr, and $T_\esc \ga 10^{12}$ yr.  Such a short scattering
time may seem unreasonable but is possible.  Very roughly this time is about
$(\Omega_e(m_e/m_p)^{q-1}f_{\rm turb})^{-1}$, where $\Omega_e\sim
20(B_\perp/\mug)$ is the electron gyrofrequency, $q$ is the spectral index of
turbulence energy density, and $f_{\rm turb}$ is the ratio of the total
turbulence energy density to the energy density of the magnetic field.  A
Kolmogorov index of 5/3 and $f_{\rm turb}\sim 10^{-6}$ would give a
$\tau_\scat\sim$ few years but a steeper spectrum will be less efficient and may
require unreasonably large energy density for the turbulence (see also below).
Figure 2 shows a comparison of these times with the total loss and crossing 
times from Figure 1.

\begin{figure}[htbp]
\leavevmode\centering
\epsscale{1.0}
\plotone{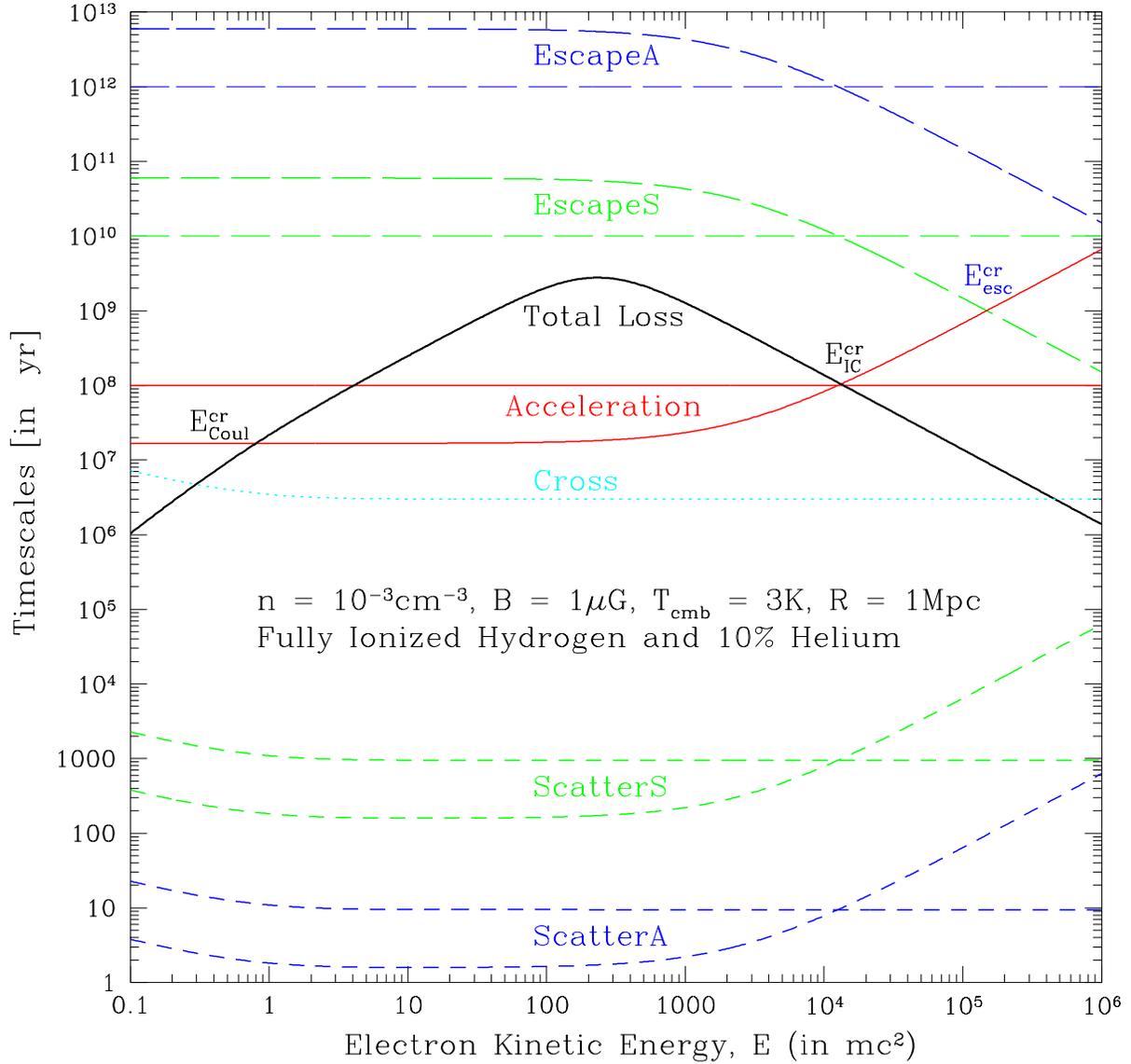}
\caption{Comparison of the energy dependence of the {\bf Total Loss} time (heavy 
solid line, from Fig. 1) with timescales for scattering (dashed), acceleration 
(solid) and escape (dot-dashed)
of electrons due to stochastic (A) and shock (S) acceleration for typical ICM 
conditions. Two examples are given. One with constant acceleration and other 
timescales, corresponding to an acceleration rate $A(E)\propto E$, and a 
second with  variable timescales where $A(E) \rightarrow$ a constant at high 
$E$, corresponding to 
an exponent $q = q'= 1$. The dotted line show 
the average crossing time $T_\tr \sim R/(c\beta)$ across a region of size 
$R\sim 1$Mpc. The critical energies where the $E$ dependent acceleration time is 
equal to the escape time, and the Coulomb and IC loss times are shown. 
}
\end{figure}

The situation is very similar for {\bf acceleration by shocks}, a first order
Fermi process, which also requires turbulence for scattering the electrons back
and forth across the shock (see e.g.  Jones 1994).  In this case we have the
additional systematic acceleration term $A_{sh}(E)\sim (\beta_{sh}/\beta)^2
D_{\mu \mu}$, where the shock velocity $c\beta_{sh}$ is of the order of the
sound velocity.  For ICM conditions $\beta_{sh}\sim 3\times 10^{-3}$ so that
shock acceleration is about hundred times faster than stochastic acceleration.
Hence, we require a 100 times longer scattering time, which requires
correspondingly smaller density of turbulence.  This and the corresponding
escape time are also sketched in Figure 2.

We should, however, note that in general these time scales are energy dependent.
For the cases discussed above one expects these time to vary as $E^{2-q}$ (see
Pryadko \& Petrosian 1997).  An example of $E$ dependent time scales (with 
$q=1$)
are also shown in Figure 2.  Note that with increasing scattering time the
distance diffused by electrons increases and reduces the above mentioned
difficulty with the spatial smoothness of the acceleration process. However, for 
a complete removal of this difficulty we need $q=1$ and $\tau_\scat \propto 
E^2$.

We now consider several scenarios with opposing and somewhat extreme
assumptions.

\subsection{Continuous Acceleration And Steady State Models}

The age of the GHz radio emitting electrons $(E\sim 10^4$) could be as low as
$10^8$ years (see Figs.  1 or 2) so that unless the observed nonthermal emission
is a short transient event of comparable time scale we require a continuous
acceleration or injection of nonthermal electrons in the ICM.  In this case
$Q(E)$ is a constant independent of time and if the density and magnetic field
change slowly, say on a Hubble time scale like the CMB photons, then on this and
shorter time scales we will be dealing with a time independent or steady state
situation with ${\partial f}/{\partial t} = 0$.  Then $f(E)$ obtained from
equation (\ref{KEQ}) represents the radiating electrons and must conform to
equation (\ref{rade}).

\subsubsection{Acceleration of Thermal Electrons}

The most likely source for the accelerated electrons might appear to be the
background hot plasma, $Q(E)=(\sqrt{\pi}/2)nE_{th}^{-3/2}\sqrt{E}
e^{-E/E_{th}}$, where $E_{th}=kT/m_ec^2=0.02$.  However, this possibility
suffers from two serious difficulties.  The first has to do with the
acceleration process.  Although acceleration by plasma turbulence of low energy
(nonrelativistic) electrons is possible (Hamilton \& Petrosian 1992), the
required conditions for it is not the case in the ICM.  Presence
of short wave (or high $k$ vector, $k=2\pi \nu_B/c\sim 6\times 10^{-10}(B/\mug)$
cm$^{-1}$) turbulence and a ratio of plasma to gyrofrequency of less than one
(or Alfv\'en velocity $\beta_A>(m_e/m_p)^{1/2}\simeq 0.023$) is required (see 
Pryadko \&
Petrosian 1997).  In the ICM $\beta_A\simeq 3\times 10^{-4}$ and the value of
this ratio is $100(n/10^{-3} \cc)^{1/2}(\mug/B)\gg 1$.  Furthermore, it is not
clear how such waves can be excited, and even if excited they will be damped
quickly because of the high temperature of the ICM (Pryadko \& Petrosian 1998,
1999).

The second and more serious difficulty in accelerating the background plasma
electrons has to do with the high Coulomb losses (already encounterd in \S {\bf 
3.1}.  The acceleration process must
overcome the heavy losses the electrons will suffer as they are pulled from
their low energy state across the energy range 10 keV to several 100 MeVs.  In
addition, for a reasonable acceleration time scale, $\tau_\ac\sim 10^8$yr, the
accelerated electron spectrum will extend into the nonrelativistic region with a
relatively steep upturn at $E\leq 0.5$, where $\tau_\Coul<\tau_\ac$ (see
Hamilton \& Petrosian 1992, Park, Petrosian \& Schwartz 1997).  This does not
agree with the desired equation (\ref{rade}), requires a high level of
turbulence ($\sim 10^{48}$ reg/s, see Blasi 2000a), and will lead to the input 
of
a high amount of energy in the ICM as in the NTB model. As
discussed in \S {\bf 3.1}, this will heat up the ICM plasma to above $10^8$ K in 
less than $10^8$ yr.

{\it We, therefore, can conclude that the background thermal 
electrons cannot be the source for the nonthermal electrons, except for a short 
period of less than $10^8$ yr}.

\subsubsection{Acceleration of Injected Non Thermal Electron} 

To overcome both of the above difficulties we require injection of relativistic
electrons, presumably from the cluster galaxies, as the initial source.  We
first consider the simplest case of a delta function injection,
$Q(E)=Q_0\delta(E-E_0)$.  The acceleration process will distribute these
electrons above and below $E_0$ and there could be other breaks at critical
energies $E_\Coul ^{cr}, E_\ic ^{cr}$ and $E_\esc ^{cr}$ where the acceleration
time $\tau_\ac=\tau_\Coul, \tau_\ic$ and $T_\esc$, respectively.  Example of
these energies are shown in Figure 2.  For a detailed discussion the reader is
referred to Petrosian (1994) and Parks \& Petrosian (1995), and for some 
examples
of complex spectra to Petrosian \& Donaghy (1999).  Here we describe some of the
possibilities relevant to the problem at hand.

Although in certain circumstances the resultant spectrum can be approximated by
a power law, this is the exception rather than the rule.  A power law spectrum
over a wide range of energies is achieved for simple diffusion coefficients and
for negligible loses.  For example, for the simple case of
\beq\label{coef}
D(E)={\cal D}E^{q'}, A(E)=a{\cal D}E^{q'-1},\,\,\, {\rm 
and}\,\,\,T_\esc=E^s/(\theta {\cal D})
\eeq 
and for the special case of $s=2-q'$ one gets 
\beq\label{spec1}
N(E) \propto Q_0\cases{(E/E_0)^{a-x+\sqrt(x^2+\theta)} &if $E<E_0$,\cr 
(E/E_0)^{a-x-\sqrt(x^2+\theta)} &if $E>E_0$,\cr}
\eeq
where $x=(a-1+q')/2$.  Here we follow the notations in Pryadko \& Petrosian
(1997) and Petrosian \& Donaghy (1999), rather than that of Park \& Petrosian
(1995) and Park et al. (1997) who use $q$ for our $q'$ here.  In what follows we 
assume $E_0<200$ MeV
and concentrate on the spectrum above $E_0$.

For $\theta\gg a \sim 1$, above $E_0$, $N\propto E^{-\sqrt(\theta)}$.  This is
the kind of acceleration model used by Schlikeiser et al.  (1987).  If $s\neq
2-q'$ the spectrum will deviate from a power law (exponentially as in modified
Bessel functions $I_n$ and $K_n$) at the energy $E_\esc^{cr}\sim \theta^{{1}
\over {|s-2+q'|}}$.  This may appear as a good explanation for the exponential
cut off needed in equation (\ref{rade}) if $E_\esc^{cr} \sim 10^4$.  However, 
this
would require $ T_\esc \ll \tau_\ic$ below this energy, which as stressed above 
is ruled out by
observations and  arguments based on energy budget.  As shown in Figure 2 for 
the two
acceleration models $\theta\sim \tau_\ac/T_\esc \sim 10^{-2}$ or $10^{-4}$ so
that $E_\esc^{cr}\gg 10^4$.  In addition, because acceleration by shocks (if
these exist in the ICM) is more efficient than by turbulence, the ratio of the
systematic acceleration rate to the diffusion rate $a = (\tau_{\rm
diff}/\tau_\ac) \sim\beta_s/\beta_A\sim 10^2$.  In this case, {\it i.e.}  in the
limit $\theta \rightarrow 0$, equation (\ref{spec1}) reduces to
\beq\label{spec2}
N(E) \propto Q_0\cases{(E/E_0)^a &if $E<E_0$,\cr 
(E/E_0)^{-q'+1} &if $E>E_0$,\cr}
\eeq
so that to obtain the index $p=3$ required in equation (\ref{rade}) we need
$q'=4$.  In general $q'$ is related to the spectral index $q$ describing the
distribution of the wave vector of the turbulence.  For Alfv\'en waves $q'=q$ so
that we require an spectrum of turbulence which is much steeper than the
commonly assumed value of 5/3 expected for a Kolmogorov spectrum.  As described
in \S {\bf 4.1} a high value of $q$ will require a high level of turbulence
especially for the stochastic acceleration model.

A more reasonable explanation of the required exponential cut off comes from
inclusion of the losses in equation (\ref{KEQ}).  As mentioned above deviation
from a power law is expected at energy $E^{cr}_\loss$ where a specific loss time
is equal to the acceleration time.  The deviation occurs in the side where loss
time is shorter.  If $\tau_\loss<\tau_{ac}$ for $E>E^{cr}_\loss$ (assume
to be $>E_0$), then the spectrum decreases sharply (approximately exponentially) 
above this energy.
This situation can arise from IC and synchrotron losses, if the acceleration
time decrease more slowly than the loss time ($\propto E^{-1}$) as in the two
examples shown in Figure 2.  This requires a systematic acceleration rate of
$A(E)\propto E^{<2}$, which for scattering by Alfv\'en waves (either in the
stochastic or shock acceleration case) implies $q=q'<3$.  As evident from
equation (\ref{spec2}) this would give rise to an accelerated electron spectral
index $p<2$ which is too small. (For the constant and variable acceleration
time scales shown in Figure 2, $q'= 2$ and 1, and $p=1$ and 0, respectively.)  
In the opposite case
of $q'>3$, the losstime $\tau_\loss<\tau_{ac}$ below $E^{cr}_\loss$ and the
spectrum steepens (becomes softer).  This situation clearly cannot produce the
exponential cut off at high energies and may arise due to Coulomb losses at low
(perhaps nonrelativistic) energies $E<E_\Coul^{cr}$.  For a thorough discussion
of all possibilities see Park \& Petrosian (1995).

Some of the above discussion is based on analytic solutions which are obtained
for simple diffusion coefficients.  In general these coefficients are more
complex (Dung \& Petrosian 1994, Pryadko \& Petrosian 1997) and the resultant
electron spectra could have other features such as a plateau just before the
exponential cut off (Park \& Petrosian 1995, Petrosian \& Donaghy 1999).
Testing these more realistic models is beyond the scope of this paper and not
warranted by the existing observations.

In summary, we can conclude that an exponential spectral cut off can be produced
at $E\sim 10^4$ if above this energy either $T_\esc < \tau_\loss$, or $T_\esc >
\tau_\loss$ but $\tau_\ac < \tau_\loss$ .  The first possibility is ruled out by
observations and the second will give rise to a flat electron spectrum with
$p\sim 1$.  Although the existing radio, HXR and EUV data do not have sufficient
spectral resolution to rule out this model, a value of $p=1$ is barley
acceptable.  Such a flat spectrum will also exacerbate the problem of low
required value for $B$.  More importantly, if the EUV radiation is also due to
the IC process, then the implied photon spectral index of $\alpha=(p+1)/2=1$
would mean a HXR to EUV flux ratio of (80-20)/(0.4-0.07)=200 which is more than
two orders of magnitude larger than the observed value of less than 2.  We note,
however, that one can specify a contrived and unphysical energy dependence of
the acceleration rate which can steepen the spectrum below a GeV to produce more
EUV photons.  We will not discuss such possibilities.

The above difficulty can not be circumvented even if the injected electron 
spectrum,
instead of being narrow as a delta function, is a broad power law; $Q(E)\propto
E^{-p_0}$ for $E>E_\min$.  In this case the final spectrum is obtained by the
convolution of $Q(E' - E)$ with the above spectra.  If we use the model of
equation (\ref{spec2}), for $p_0>q'-1$ this convolution will have no effect
above $E_\min$ and the difficulty remains.  But in the opposite case,
$p_0<q'-1$, the acceleration process will have a negligible effect, and the
resultant spectrum will be essentially same as the injected spectrum which is
now even flatter.  Thus, we conclude that

\noindent
{\it the steady state acceleration in the ICM of either thermal or non 
thermal electrons can not produce the requisite spectrum for reasonable 
physical conditions.}

\subsubsection{Transport Effects and Cooling Spectra}

Considering the difficulties with the acceleration in the ICM discussed above we
now explore the possibility that electrons are accelerated somewhere else,
presumably in galaxies, and are injected into the ICM, where they undergo only
scattering and losses.  In this case we still need some kind of turbulence to
scatter and trap the electrons in the ICM, but we assume that these only
isotropize the electrons and diffuse them spatially but cause neither diffusion
in energy nor acceleration.  As before, the scattering rate determines the
escape time in equation (\ref{KEQ}) where now we set $D(E)=A(E)=0$.  Because we
are interested in relativistic electrons, we approximate the loss term in
equation (\ref{KEQ}) as
\beq\label{loss}
{\dot E}_L (E)/E_p=(1+(E/E_p)^2)/\tau_0,
\eeq
where
\beq\label{values}
\tau_0=(4\pi r_0^2cn\Clog)^{-1}=6.3\times 10^9(10^{-3}\cc/n)\,{\rm yr}\,\,\,\,\ 
{\rm and}\,\,\,\,\ 
E_p\simeq[(9/8)(n\Clog/(u_{ph}+B^2/8\pi))]^{1/2} = 235
\eeq
are approximately the loss time (multiplied by 2) and the energy where the {\bf 
Total Loss} curve reaches its maximum (see 
Figs. 1 and 2). Here we have ignored the bremsstrahlung loss and the weak 
dependence on $E$ of Coulomb losses at nonrelativistic energies. Solution of 
equation (\ref{KEQ}) will then give the effective spectrum of the
radiating electrons (referred commonly to as a {\bf cooling spectrum}) that must 
conform to equation (\ref{rade}).  $Q(E)$ represents the average rate of 
injection of accelerated electrons, which in general will have a broad 
distribution. A reasonable (and convenient) form is a power law, $Q(E)=Q_0 
(E/E_p)^{-{p_0}}$ 
with 
the same low energy constraints as those used in connection with equation 
(\ref{rade}).
Because most of 
the energy of electrons with $E<100$ MeV goes into heating the ICM, the rate 
of 
injection of energy  below this value must be less than the SXR thermal 
luminosity. Therefore, as stated in the previous section the spectrum of the 
injected electrons must drop off sharply below 8 MeV or have an spectral index 
$p_l<0.5$. It should also be noted that the sources of injection must be 
sufficiently numerous and 
have a distribution such that they can produce a surface brightness distribution 
that is as smooth as that observed at radio wavelengths. 

It can then be shown  that for a 
finite, and perhaps energy dependent, $T_\esc$ the steady state solution of 
equation (\ref{KEQ}) is
\beq\label{escape}
N(E)={\dot E}_L^{-1}e^{-x}\int_E^\infty Q(E)e^xdE,\,\,\,\ {\rm with}\,\,\,\  
dx=-dE/(T_\esc{\dot E}_L). 
\eeq
This is a partially cooled spectrum and has a break at $x\sim 1$ or at energy
$E_{cr}$ where $T_\esc=\tau_\loss$.  For $x\ll 1$ or $T_\esc\gg \tau_\loss$ one
expect a fully cooled spectrum and for the opposite limit, $T_\esc\ll
\tau_\loss$, the spectrum is same as the injected spectrum multiplied by
$T_\esc$.  For example, for $T_\esc={\cal T_\esc}(E/E_p)^{\nu -1}$ and for
energies above the maximum of the $\tau_\loss$ curve at about 100 MeV, where
$\tau_\loss\propto E^{-1}$, a power law injected spectrum (for $\nu>0$ and 
$p_0>1$) gives
\beq\label{inside}
N(E) = Q_0 \cases{\tau_0+(E/E_p)^{-p_0-1}/\nu &if  
$E_{cr}\ll E$, \cr
{\cal T}_\esc(E/E_p)^{-p_0-1+\nu} &if  $E\ll E_{cr}$,\cr}
\eeq
where $E_{cr}=E_p(\nu {\cal T}_\esc/\tau_0)^{-1/\nu}$.  Thus, for
$p_0\sim 3$ and $\nu\sim 0$ and $T_\esc\simeq 0.02\tau_0$ we  obtain a spectrum 
with a break at $E\sim 10^4$, in good agreement with the radio
data (Rephaeli 1979 model).  However, a large fraction of the $E<E_p$ electrons 
escape from the ICM,
or more accurately from the turbulent confining region with a flux of $F_\esc(E)
\propto N(E)/T_\esc(E)$.  As already pointed out above this is in
disagreement with the observations.  This difficulty is even more severe for a
narrow injected spectrum, {\it e.g.}, a delta function.

For the more reasonable case of $T_\esc\gg\tau_\loss$ equation (\ref{escape})
reduces to the fully cooled spectrum of $N(E)={\dot E_\loss}^{-1}\int_E^\infty
Q(E)dE$.  For a delta function injection at a high energy the spectrum of the
radiating electrons will vary as ${\dot E}_\loss^{-1}$ which will be essentially
constant up to 200 MeV and then decrease with a power law index $p=2$.  This
does not agree with equation (\ref{rade}).  For a power law injected spectrum
$N(E)=(Q_0/(p_0-1))(E/E_p)^{-p_0}\tau_\loss$, where $\tau_\loss$ is given by the
heavy solid line in Figure 2.  For $p_0=2$ this will agree roughly with the data
but not with the more accurate model of equation (\ref{rade}) with a break at
$E\sim 10^4$.  One way to have such a feature is if the injected electrons
obtain the imprint of the break at their sources.  In this case by substitution 
of
equation (\ref{rade}) for $N$ in equation (\ref{KEQ}) with $D=A=T_\esc^{-1}=0$
we find the necessary injected steady state spectrum to be $Q(E) \propto
E^{2-p}e^{(-E/E_{cr})}[(2-p)/E - 1]$ at high energies and with a similar
expression at lower energies.  This of course is an {\it ad hoc} assumption and
does not clarify the acceleration mechanism.

Thus, unless there 
exists  an arbitrary and contrived injected spectrum,  we must conclude that 

\noindent
{\it the steady state injection and cooling model also fails to 
describe the observations adequately}. 

\subsection{Time Dependent Models}

The upshot of the discussion in the previous section is that a steady state
acceleration in the ICM or modification of a simple accelerated spectra by
transport processes in the ICM fail to reproduce the general features of the
required spectrum.  We therefore consider time dependent scenarios with time
variation shorter than the Hubble time.  In this case we consider solutions of
the time dependent equation (\ref{KEQ}).  We start with the generic model of a
prompt one time injection of electrons with $Q(E,t)=Q(E)\delta(t-t_0)$.  More
complex temporal behaviors can be obtained by the convolution of the injection
time profile with the solutions described below. Similar, but somewhat 
different, treatments of the following cases can be found in Sarazin (1999) and 
Beunetti et al. (2000)

\subsubsection{Transport Effects}

We first consider the transport effects in the ICM without any acceleration.  In
this case the time-dependent equation (\ref{KEQ}), with $D=A=0$, $T_\esc$
independent of time and energy, and ${\dot E}_L$ constant in time, has the
following formal solution:
\beq\label{prompt}
f(E,t) = \exp\{-t/T_\esc\}Q(E'(E,t)){\dot E}_L(E'(E,t))/{\dot E}_L(E),
\eeq
where $E'(E,t)=\tau^{\rm inv}(\tau(E)-t)$ and $\tau^{\rm inv}$ is the inverse 
function of 
\beq\label{tau}
\tau(E)=\int_E^\infty dE/{\dot E}_L(E).
\eeq
Using equation (\ref{loss}) for ${\dot E}_L(E)$, we find $\tau(E)/\tau_0=\pi/2 -
\tan^{-1}(E/E_p)$, $\tau^{\rm inv}(x)=\coth x$ and
$E'/E_p=(E/E_p+\tan(t/\tau_0))/(1 - (E/E_p)\tan(t/\tau_0))$, so that for a power
law injection, $Q(E)=Q_0(E/E_p)^{-p_0}$, the solution for $p_0\geq 2$ becomes
\beq\label{prompt1}
f(E,t) = \exp \{-t/T_\esc\}Q_0{{[1 - (E/E_p)\tan(t/\tau_0)]^{p_0-2}} \over 
{\cos^2(t/\tau_0)[E/E_p+\tan(t/\tau_0)]^{p_0}}}.
\eeq
The solid lines in Figure 3 shows the spectral evolution according to this 
equation for
the specified parameters and for several times past the injection epoch.  At
early times the spectrum is a power law in the energy range
$\tan(t/\tau_0)<E/E_p<1/\tan(t/\tau_0)$, goes to zero at
$E/E_p=1/\tan(t/\tau_0)$, and is flat for $E/E_p<\tan(t/\tau_0)$.  As expected
the power law extends to $E\ga 4\times 10^4$ needed for radio production only
for a short period of $\tau_0/100$ {\it i.e.} $\la 10^8$yr.  The power law 
portion disappears
for $t>\pi \tau_0/4\sim 5\times 10^9$yr and we obtain a degenerate flat spectrum
extending to $E_p$.  Between this and $t>\pi \tau_0/2\sim 10^{10}$yr the cut off
moves to lower energies and the amplitude drops as $\tan^{p_0-2}(t/\tau_0)$.  In
addition, the spectrum decays exponentially on a time scale of $T_\esc$ so that
we need $T_\esc> \tau_0$ which requires presence of turbulence or chaotic
fields.  Such a turbulence can also accelerate the electrons.  Therefore, the
above spectra are correct if the acceleration time is longer than $\tau_0$.

It is, therefore, clear that either the observable duration of the nonthermal
activity in the clusters is a rare phenomenon or we need episodic injection of
electrons on a timescale of 100 million years.  Whether mergers and resulting
shocks, or AGN activities can provide such a source is unknown.  If this is the
case then the rapid cut off at $E/E_p=1/\tan(t/tau_0)$ may mimic the exponential
form of equation (\ref{rade}), so that with $p_0 \lesssim 3$ this model will be
acceptable.

\begin{figure}[htbp]
\leavevmode\centering
\epsscale{1.0}
\plotone{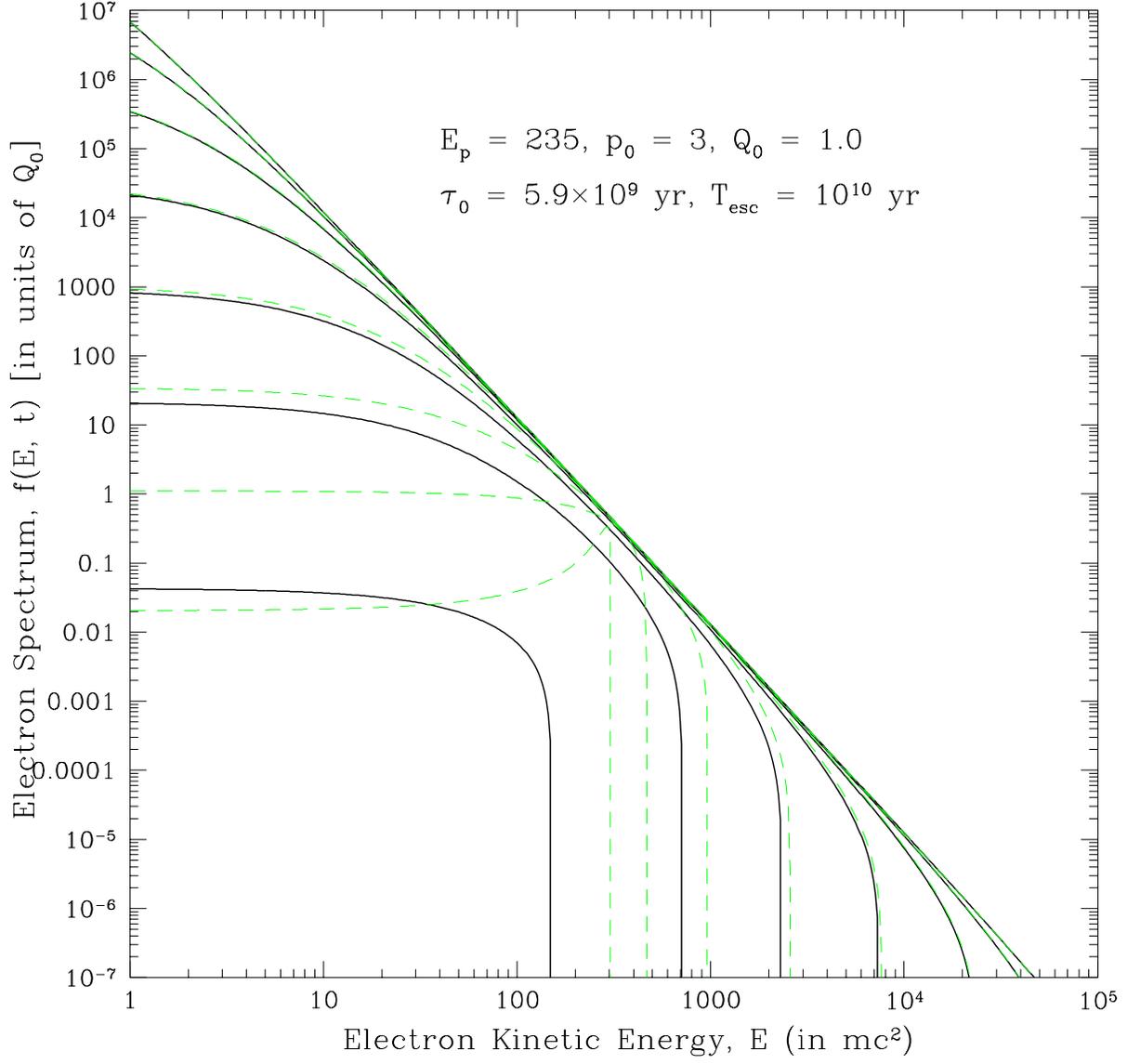}
\caption{Evolution with time of a power law injected spectrum (top line) 
subject to Coulomb and IC (plus synchrotron) losses as given by equation 
(\ref{prompt1}) (solid lines for times $t_n=10^{n/2}\tau_0, n=-6$ to 0) and with 
acceleration ($b=2, \delta=0$) obtained from equation (\ref{prompt3}) (dashed 
lines for times $t_n=10^{n/2}\tau_0, n=-6$ to 1).}
\end{figure}

\subsubsection{Acceleration Plus Transport}

A more varied and complex set of spectra can be obtained if we add the effects
of diffusion and acceleration.  Simple analytic solutions for the time dependent
case are possible only for special cases.  Most of the difficulty arises because
of the diffusion term which plays a vital role in shaping the spectrum for a
narrow injection spectrum.  For some examples see Park \& Petrosian (1996).  As
we have seen for the steady state case the effect of the diffusion is important
for a narrow injected spectrum.  Here we will limit our discussion to a broad
initial electron spectrum in which case the effects of this term can be ignored.
Thus, if we set $D(E)=0$, then the solution (\ref{prompt}) of equation
{\ref{KEQ}) can be generalized simply by inclusion of the systematic
acceleration term $A(E)$ in ${\dot E}_L$ (see eq.  [\ref{coef}] and 
[\ref{loss}]) as
\beq\label{acandloss} 
{\dot E}_L (E)/E_p=(1+(E/E_p)^2 - b(E/E_p)^{q'-1})/\tau_0,
\eeq 
where $b=a{\cal D}\tau_0 E_p^{q'}=\tau_0/\tau_\ac(E_p)\sim 10^2$ or 1 for the
shock or stochastic accelerations, respectively, and for the parameters 
described in
the previous section.  For a general exponent $q'$ one must resort to numerical
solutions.  For the purpose of demonstration of the effects of further ICM
acceleration we consider the simple case of $q'=2$ (corresponding to the
constant acceleration timescale of Fig. 2), which has a solution similar to
that shown by equation (\ref{prompt1}):
\beq\label{prompt2}
f(E,t) = \exp \{-t/T_\esc\}Q_0{{[T_+ - (E/E_p)\tan(\delta
t/\tau_0)/\delta]^{p_0-2}} \over {\cos^2(\delta t/\tau_0)[
T_-(E/E_p)+\tan(\delta t/\tau_0)/\delta]^{p_0}}}, 
\eeq 
where $\delta^2 = 1-b^2/4$ and $T_{\pm} = 1 \pm b\tan(\delta
t/\tau_0)/(2\delta)$.  This solution (valid for $b^2<4$) reduces to that in
(\ref{prompt1}) for $b=0$.  For $b^2>4$ we are dealing with an imaginary value
for $\delta$ so that tangents and cosines become hyperbolic functions with
$\delta^2 =b^2/4 - 1$.  For $\delta=0$ or $b=2$ either form reduces to
\beq\label{prompt3} 
f(E,t) = \exp \{-t/T_\esc\}Q_0{{[1 - (E/E_p -
1)t/\tau_0]^{p_0-2}} \over {[E/E_p-(E/E_p - 1)t/\tau_0]^{p_0}}}.  
\eeq 
The dashed line in Figure 3 show the evolution of spectra for this latter case
and Figure 4 shows the solution according to equation (\ref{prompt2}) for larger
values of $b =5$.  As expected with acceleration one can push the electron
spectra to higher levels and extend it to higher energies, but as described
below this does not significantly alter the above conclusion based on the
transport effects alone, but improves the situation somewhat.

For $b<2$ the situation is similar to the case $b=0$ (no acceleration) except 
that the spectra decay more slowly; the degenerate phase of a flat spectrum is 
reached later and extends to a higher energy. For $b>\sqrt(2+4/p_0)\sim 1.83$
a local maximum appears during the degenerate phase just below the maximum 
energy (see Figs. 
3 and 4). This peak could be very high and narrow. For $b>2$ the acceleration 
becomes more and more important and can quickly reverse the decay and give 
rise to a growing spectrum. As before, for early times one gets a power law 
spectrum at $\tanh(\delta t/\tau_0)/(\delta T_-)<E/E_p<T_+\delta/\tanh(\delta 
t/\tau_0)$. The spectrum now can be sustained to a high energy for all 
times: 
\beq\label{Emax}
E_{\rm max}/E_p = 1 + (b/2)\tanh(\delta t/\tau_0)/\delta/\tanh(\delta 
t/\tau_0)/\delta > \delta + b/2\simeq b.
\eeq
Thus, with faster acceleration rate ({\it i.e.}  $b>50$) we can have electron
spectra extended above $10^4$ MeV.  However, the period when the spectrum below
this energy is a power law is short.  The degenerate phase is reached quickly
when $\tanh(\delta t/\tau_0) = \sqrt(\delta/ (\delta + 2))$.  For large values
of $b$ this gives $t/\tau_0=\ln(\delta+1+\sqrt((\delta+1)^2+1))/2\delta
\rightarrow \ln(2\delta)/(2\delta)$ which is less than $5\times 10^8$ yr for $b
>50$ implying a short duration for the power law phase.  As evident from Figure 
4
soon after the electrons are reaccelerated to above $10^4$ MeV the power law
portion disappears.  Of course, the situation can be improved with a more 
complex
injected spectrum ({\it e.g.}  a broken power law, see Brunetti et al.  2000) or
with a time dependent injection and/or acceleration parameters.  However, some
fine tuning may be required to sustain the required spectrum for a period
significantly longer than $10^8$ years, which is essentially determined by the 
temperature of the CMB and the resultant lifetime of the $E=10^4$ electrons. In 
any case the additional acceleration 
in the ICM improves the situation.

\begin{figure}[htbp]
\leavevmode\centering
\epsscale{1.1}
\plottwo{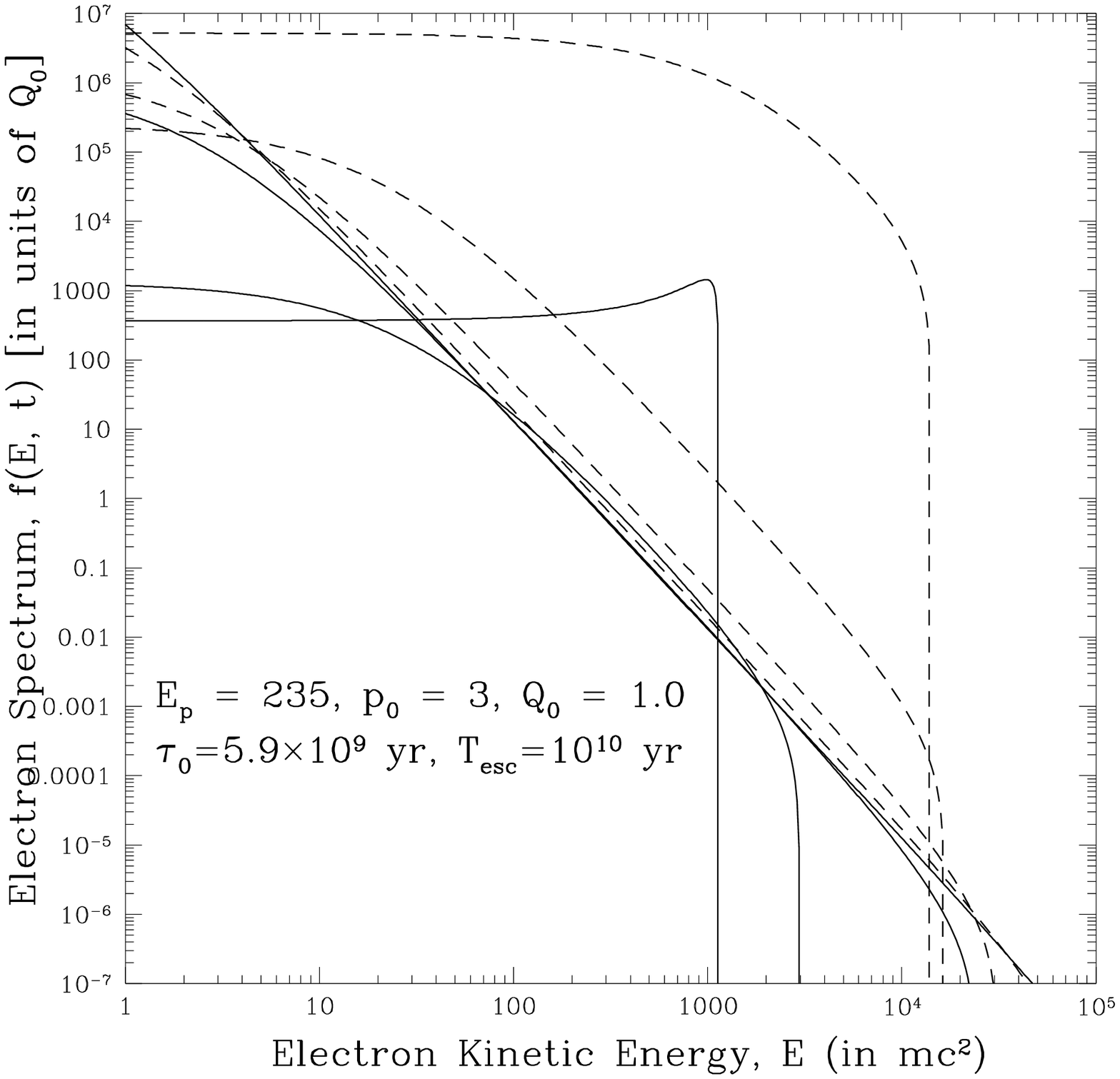}{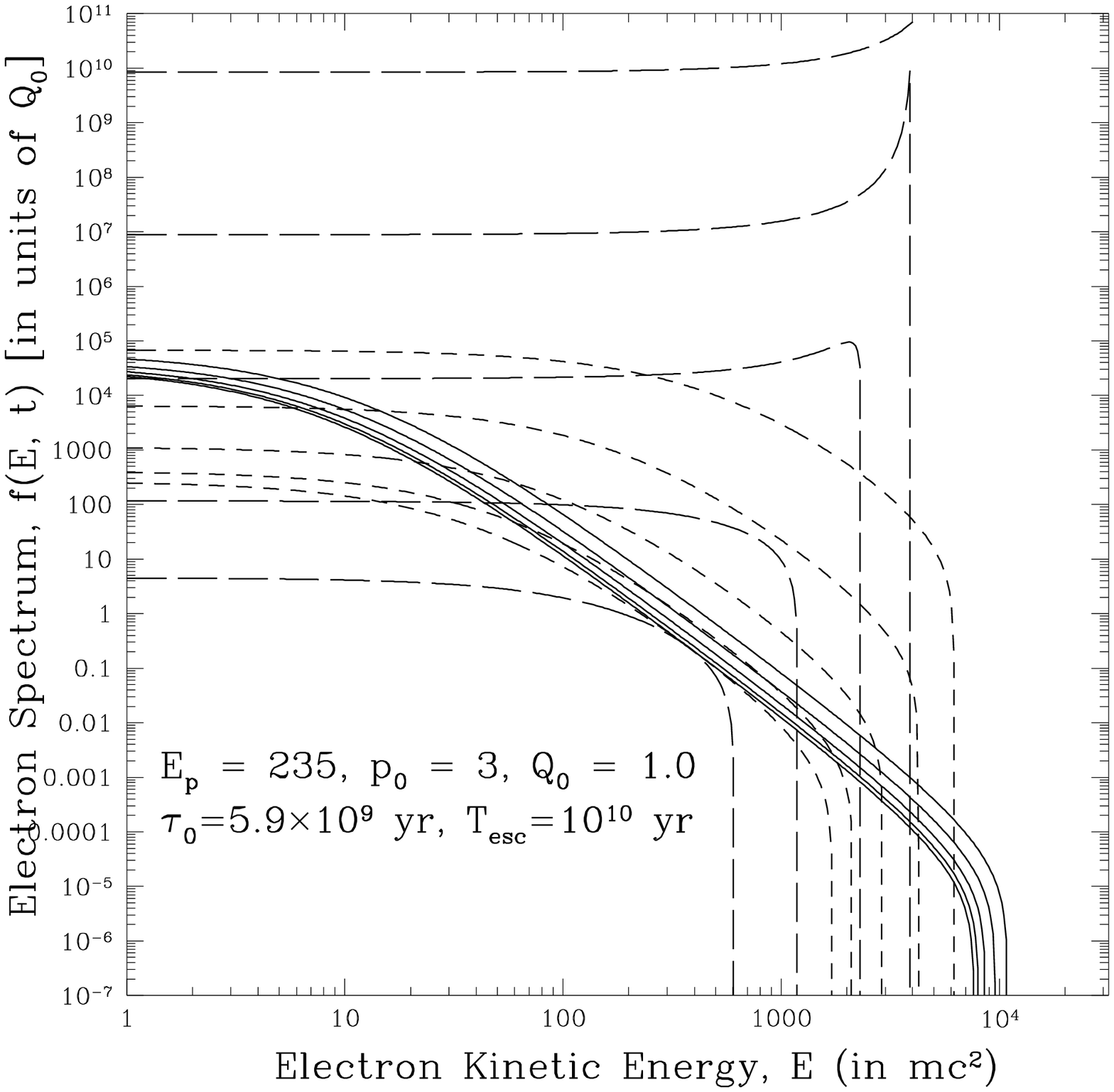}
\caption{Same as Fig. 3. {\bf Left Panel:} For $b=5$ (solid lines, for $t/\tau_0 
= 10^{-3}, 
10^{-2}, 10^{-1}$ and 1) and $b=60$ (dashed lines, for $t/\tau_0 c= 10^{-2.5}, 
10^{-2}, 10^{-1.5}$ and $10^{-1}$):
{\bf Right Panel:} For $b=2, 5, 9, 17$ and 26 (from lower to upper 
spectra) and at three different times, $t/\tau_0= 10^{-1.5},  10^{-0.8}$ and 
$10^{-0.2}$ (solid, dotted and dashed lines, respectively), obtained from 
equations (\ref{prompt2}) and (\ref{prompt3}).}
\end{figure}

\section{SUMMARY AND DISCUSSION}

The purpose of this paper has been to investigate the emission mechanisms for
the observed nonthermal radiation from the ICM of several clusters and to
explore possible acceleration scenarios.  We have used the observations of Coma 
cluster for our quantitative analyses. The qualitative aspects of the results 
summarized below are quite general, but the specific values of the parameters 
depend on the assumed values of the density, temperature, size, magnetic fiels 
etc, some of which are poorly known and can vary from cluster to cluster.

For the radiation mechanism, we have come to two important conclusions.

1) The source of the HXR flux cannot be nonthermal bremsstrahlung emission by
semi-relativistic electrons because of the extreme inefficiency of this process,
unless this is a short-lived ($<10^8$ yr) phenomenon.

2) Inverse Compton scattering of relativistic electrons by the CMB photons is a
more natural process for production of both the HXR and EUV emissions.  We have
shown that the problems with a low value of magnetic field needed for this
mechanism (discussed widely in the literature) can be alleviated when we include
the effects of more realistic (broken power low) spectra and anisotropies in the
pitch angle distribution of the electrons.  Observational selection bias can
also favor the IC emission at low magnetic fields.

Combining the requirements of the IC process for HXR and EUV emissions with the
requirements of the synchrotron process for the radio emission, we derive a
simple spectrum for the radiating electrons as described by equation
(\ref{rade}).

Next we investigate the constraints that this spectrum, and other
considerations, put on the acceleration mechanism.  We consider both second
order Fermi stochastic acceleration by turbulence and first order Fermi
acceleration by shocks.  We derive parameters for both these mechanisms so that
they can accelerate electrons to the required energies of $E>10^4$ MeV within
their life time of $10^8$ years or shorter.  The important conclusions here are
the following:

1) The ICM must contain a high level of turbulence (or other scattering agent)
to trap the electrons for time periods longer than their loss timescales and
much longer than their crossing time across the cluster.

2) Acceleration of the thermal ICM electrons to relativistic energies will be
difficult given the low value of the Alfv\'en velocity, and more importantly
requires input of a large amount of energy in the ICM.  It will also give rise
to an unacceptable spectrum for the IC model.

3) Steady state acceleration of injected relativistic electrons gives rise to a
flatter spectrum than desired, or to a HXR and EUV source that extends well
beyond the boundaries of the radio source.

4) Steady state cooling of a power law injected spectrum also suffers from the
same shortcoming or must involve {\it ad hoc} assumptions.

5) Time dependent models fair much better.  A power law injected spectrum, under
the influence of transport effects alone, can evolve into one with a high energy
cut off at $E_{cr}\sim 10^4$ (as required by the observations) after a time
equal to the energy loss time at this energy, which is about $10^8$ yr.  For
later times ($t>\tau_0\sim 6\times 10^9$ yr) the cut off moves to lower energies
and the spectrum becomes flat below it.  If one adds an acceleration agent then
the spectrum can be maintained above the desired energy for a longer period.
This requires an acceleration time scale that is shorter than $10^8$ yr.  But,
at such high acceleration rates, the spectrum below this cut off becomes flat in
a shorter period of time, $t\sim\tau_0 \ln b/b$, where $b\sim \tau_0/\tau_\ac$.
This can yield an acceptable spectrum for a period of about $5\times 10^8$ yr.

The above results mean that either the nonthermal emissions from the ICM are
short lived and rare events or there is episodic injection of power law spectrum
of relativistic electrons on a time scale of about $10^8$ yr.  This, however,
still leaves the initial mechanism of the electron acceleration unresolved.  A
likely scenario is that episodic mergers of sub clusters or encounters between
galaxies can give rise to shocks and turbulence.  The initial acceleration can
take place in these shocks.  The spectrum of radiating electrons is a result of
transport and further acceleration (by turbulence) in the ICM.  In such a
situation, however, one would expect a different spatial distribution for the
EUV emission than for HXR and radio emissions.  The latter emitted by higher
energy, shorter lived electrons will be more concentrated around the initial
source.  Similarly, a radial variation of the magnetic field could result in a 
more (or less) centrally concentrated synchrotron (radio) emission compared to 
the IC (HXR and perhaps EUV) emision. Density variations will affect mainly the 
bremsstrahlung emission relative to the other radiative processes but not the 
arguments based on the bremsstrahlung yield. Temperature variations can effect 
the spectrum of the turbulence.

An exact evaluation of the relative spatial distributions at different energy 
bands will require
solution of an inhomogeneons Fokker-Planck equation which in turn requires
knowledge of energy and spatial dependences of scattering and escape processes,
as well as spatial variation of density and magnetic field.  This is beyond the
scope of the present paper and not warranted by the existing observations.
Higher spatial resolution observation will be helpful here.

Alternative sites of the initial acceleration may be in galaxies, in which case
the homogeneous model will be a good approximation.  However, in this case, in
addition to electrons one would expect a larger energy input in form of protons.
It is likely that protons may be the source of the turbulence which is essential
for any viable model of nonthermal emission from ICM.

This work was started while I was a visitor at the Institute For Advanced 
Studies and Bochum University. I would like to thank the support of both 
Institutions, their staff and 
acknowledge stimulating discussions with Drs. Bahcall, Kumar, Schlickeiser. I 
would also like to thank Drs. Blasi, Bykov, Eilek, En\ss lin and Rephaeli for 
helpful discussions and several suggestions which led to the improvement of this 
paper.

\newpage

\end{document}